\documentstyle[12pt,psfig]{article}

\def\al{\alpha}
\def\si{\sigma}

\def\frac#1#2{{\textstyle{{#1}\over {#2}}}}

\def\lsim{\mathrel{\rlap{\lower4pt\hbox{\hskip1pt$\sim$}}
    \raise1pt\hbox{$<$}}}
\def\gsim{\mathrel{\rlap{\lower4pt\hbox{\hskip1pt$\sim$}}
    \raise1pt\hbox{$>$}}}
\def\sqr#1#2{{\vcenter{\vbox{\hrule height.#2pt
         \hbox{\vrule width.#2pt height#1pt \kern#1pt
         \vrule width.#2pt}
         \hrule height.#2pt}}}}

\newcommand{\beq}{\begin{equation}}
\newcommand{\eeq}{\end{equation}}
\newcommand{\bea}{\begin{eqnarray}}
\newcommand{\eea}{\end{eqnarray}}

\renewenvironment{thebibliography}[1]
 { \rm
   \begin{list}{\arabic{enumi}.}
    {\usecounter{enumi} \setlength{\parsep}{0pt}
     \setlength{\itemsep}{3pt} \settowidth{\labelwidth}{#1.}
     \sloppy
    }}{\end{list}}
 
\begin{document}
\titlepage
 
\begin{flushright}
{UAlg/TP/98-3\\}
{July 1999\\}
\end{flushright}
\vglue 1cm
	    
\begin{center}
{\bf Sphaleron Transition Rate in Presence of Dynamical Fermions \\}
\end{center}
\vglue 1.0cm
\begin{table}[hb]
\centerline{\begin{tabular}{ccc}
A.\ Kovner & $\ \ \ \ \ $ & A.\ Krasnitz, R.\ Potting \\
$\ \ $ & & $\ \ $ \\
{\it Theoretical Physics} & & {\it Universidade do Algarve} \\
{\it Oxford University} & & {\it UCEH} \\
{\it 1 Keble Road} & & {\it Campus de Gambelas} \\
{\it Oxford OX1 3NP} & & {\it P-8000 Faro} \\
{\it UK} & & {\it Portugal} \\
\end{tabular}}
\end{table}

\vglue 1cm 
{\rightskip=3pc\leftskip=3pc
\noindent We investigate the effect of dynamical fermions on the sphaleron
transition rate at finite temperature for the Abelian Higgs model in
one spatial dimension.  The fermion degrees of freedom are included
through bosonization.  Using a numerical simulation, we find that
massless fermions do not change the rate within the measurement
accuracy.  Surprisingly, the exponential dependence of the sphaleron
energy on the Yukawa coupling is not borne out by the transition rate,
which shows a very weak dependence on the fermion mass.
}
 
\newpage

\section{Introduction.}

The 1+1 dimensional Abelian Higgs model merits interest for physical properties
it shares with the electroweak theory. In particular, it has topologically
distinct minima of the energy, corresponding to different winding numbers of
the scalar field.
Transitions between these states are possible, at zero temperature
through quantum tunneling, and at finite temperature also by thermal activation.
What makes these transitions physically
very interesting is that they are accompanied by an anomalous change in
the fermion number,
if the model includes a chiral coupling of the gauge field to fermions.
In the electroweak theory, these transitions were, in all probability,
responsible for the erasure of the primordial
baryon asymmetry. Moreover, they may have led to electroweak baryogenesis.
Processes of these type are called sphaleron transitions, owing their name to
sphalerons, the lowest-barrier configurations separating energy minima.

Since processes violating the fermion number involve field configurations
which are nonperturbatively far from the trivial vacuum, the problem requires a
nonperturbative treatment. A useful nonperturbative framework is provided
by the Euclidean
lattice field theory, wherever processes at zero temperature or
static thermal properties are concerned.
However, the fermion-number violating processes in question
occur in real time and at a finite temperature, and thus are out of reach
for the Euclidean quantum theory. The problem simplifies considerably only
in the classical approximation. Real-time thermal properties of the
resulting classical field theory can be studied numerically after lattice
discretization. It has been established recently that, under certain
conditions, the classical approximation is reliable for the processes in
question \cite{smaa}.

There is by now an extensive body of work devoted to
numerical study of the sphaleron
transition rate in the classical approximation.
Over time, the attention has shifted from one-dimensional
models \cite{grs,bf,kp,fkp,st} to realistic aproximations of the electroweak 
theory in 3+1 dimensions \cite{aaps,ak31,gm,gmt}.
Nevertheless, the one-dimensional models have not yet exhausted their
utility. In particular, they can be used to investigate the role
of dynamical fermions, such as those present
in the standard model, in the real-time processes of
interest. While in 3+1 dimensions these degrees of freedom resist
classical treatment, such treatment is possible in one-dimensional
models upon bosonization. This approach was first proposed by 
Roberge \cite{roberge90} who used it to study static properties
of sphalerons at a finite fermion density \cite{roberge94}.
Here we apply this approach to study the real-time
dynamical evolution of the Abelian Higgs model coupled to fermions.

The effect of fermions on sphalerons has been investigated
in a variety of ways. These include perturbation theory \cite{moore,mitya},
valence approximation \cite{kuntz}, and expansion in the number of fermion 
families \cite{gr}. These methods are used to determine the
sphaleron (free) energy rather than the sphaleron transition rate, a dynamical
quantity whose determination requires a real-time treatment at finite 
temperature. Recently Aarts and Smit \cite{smaaf} used the expansion in the 
large number of
families for a numerical study of fermions in a classical Bose field background.
The latter method is costly numerically and is yet to yield a figure for the 
sphaleron rate.
To the best of our knowledge, the current work is the first calculation
of the sphaleron rate to date to account for dynamical fermions.

The contents of the paper is as follows.
In Section 2 we discuss the bosonized 
form of the model and its vacuum structure.
In Section 3 we determine the 
variation of the sphaleron energy with the Yukawa coupling.
Our numerical 
results for the sphaleron transition rate are presented in Section 4.
Section 5 contains the discussion.

\section{The model}

Our starting point is the Lagrangian density in two-dimensional space-time
\bea
{\cal L}&=&\bar{\psi} i\gamma^\mu(\partial_\mu - ie\gamma^5 A_\mu)\psi
   - \frac{1}{4} F_{\mu\nu}F^{\mu\nu} + {1\over 2}(D_\mu\phi)(D^\mu\phi)^*
 -{\lambda\over 4}(|\phi|^2- v^2)^2 \nonumber\\
&&{}-y\,[\phi(\bar\psi\psi+i\bar\psi\gamma_5\psi)
+\phi^*(\bar\psi\psi-i\bar\psi\gamma_5\psi)],
\label{eq:lagr}
\eea
using the standard notation for the two-component spinor fermions $\psi$,
the complex scalar field $\phi$, and the U(1) gauge field $A_\mu$. Here 
$D_\mu = \partial_\mu -2ieA_\mu$, where $e$ is the gauge coupling. 
We assume the fields to satisfy periodic boundary conditions. The 
scalar self-coupling and the Yukawa coupling are $\lambda$ and $y$, 
respectively.  This model is to be regulated such that the gauged current,
$\bar{\psi}\gamma^\mu\gamma^5 \psi$
is conserved while the vector current obeys the anomaly equation
\beq
   \partial_\mu\, \bar{\psi}\gamma^\mu\psi\equiv \partial_\mu J^\mu =
    -\frac{e}{2\pi} \epsilon^{\mu\nu} F_{\mu\nu}.
\label{eq:anomaly}
\eeq
In its global form,
the anomaly equation means that the variation of the baryon number $B$ 
equals that of Chern-Simons number $N_{\rm CS}$:
\beq
{d\over{dt}}(B-N_{\rm CS})={d\over{dt}}\left(\int{\rm d}x\psi^\dagger\psi
+{e\over\pi}\int{\rm d}x A_1\right)=0.
\eeq
Better suited for our purposes is the Bose-equivalent form of the Lagrangian:
\bea
{\cal L}&=&\frac{1}{2}\left(\partial_\mu \si
          -  \frac{e\sqrt\hbar}{\sqrt\pi} A_\mu\right)^2
   - \frac{1}{4} F_{\mu\nu}F^{\mu\nu} + {1\over 2}(D_\mu\phi)(D^\mu\phi)^*
  -{\lambda\over 4}(|\phi|^2-v^2)^2\nonumber\\
   &&{}+Y \left(\phi e^{-2i\sqrt{\pi/\hbar}\si}
              + \phi^* e^{2i\sqrt{\pi/\hbar}\si}\right)
 \label{eq:Lagrangian}
\eea
This form is obtained by introducing a real scalar field $\sigma$ related to
the fermion currents
via\footnote{These formulae are slightly different from the standard
bosonization formulae which do not contain the vector potential in the
bosonized expressions for the currents. The reason for this difference
is that in our model the fermions couple to the gauge field through
the axial rather than the vector coupling. The Wilson line factor which must
be included in the gauge invariant regularized expressions for the
fermionic currents are therefore different here from the standard one.
The Wilson line in the local limit does not reduce to one 
and is responsible for the appearance of the vector potential on the
right hand side of eq.(\ref{eq:curr}).}
\bea
\label{eq:curr}
\bar\psi
\gamma_\mu\psi={\sqrt\hbar\over\sqrt\pi}\epsilon_{\mu\nu}[\partial_\nu\sigma
-{e\sqrt\hbar\over\sqrt\pi}A_\nu]\nonumber\\
\bar\psi\gamma_\mu\gamma_5\psi={\sqrt\hbar\over\sqrt\pi}[\partial_\mu\sigma
-{e\sqrt\hbar\over\sqrt\pi}A_\mu]
\eea
The vector current obviously
obeys the anomaly equation (\ref{eq:anomaly}).

Two comments are in order with regard to (\ref{eq:Lagrangian}). First, note 
that, in order to satisfy periodic boundary conditions for 
(\ref{eq:Lagrangian}), the field $\sigma$ only needs to be periodic modulo
$\sqrt{\hbar\pi}$. Similarly to the winding number of the scalar field,
the periodicity mismatch of $\sigma$, 
$\int{\rm d}x\partial_x\sigma/\sqrt{\hbar\pi}$, changes by an integer under
topologically nontrivial gauge transformations. However, unlike the winding
number of $\phi$, the periodicity mismatch of $\sigma$ cannot be changed
dynamically, because the time derivative of $\sigma$ 
does satisfy periodic boundary conditions.
Thus, imposing periodic boundary conditions on $\sigma$ is a matter of a gauge 
choice. We will assume $\sigma$ to be periodic in space when we solve the model
numerically.

Secondly, note that we have included explicitly Planck's constant, obviating
the fact that the bosonization essentially links
{\it quantum} theories. In order to determine the loop expansion
parameter for the model, we re-express the fields in units of $v$.
For convenience, we also express the
coordinates in units $1/v\sqrt{\lambda}$. We then obtain for the Lagrangian
\bea
{{\cal L}\over{\lambda v^4}}&=&\frac{1}{2}\left(\partial_\mu \si
          -  g\sqrt{\hbar\over{\pi v^2}} A_\mu\right)^2
   - \frac{1}{4} F_{\mu\nu}F^{\mu\nu} + (D_\mu\phi)(D^\mu\phi)^*
  -{1\over 4}(|\phi|^2-1)^2\nonumber\\
   &&{}+{\cal Y} \left(\phi e^{-2i\sqrt{\pi v^2/\hbar}\si}
              + \phi^* e^{2i\sqrt{\pi v^2/\hbar}\si}\right),
 \label{lagresc}
\eea
where now $D_\mu=\partial_\mu-2igA_\mu$, $g=e/\sqrt{\lambda}$ and 
${\cal Y}=Y/\lambda v^3$.
It is now evident that the loop expansion parameter for the model is
$\hbar/v^2$. This parameter must be small for the classical 
approximation to make sense. In the absence of fermions $v^2/\hbar$ is an
overall factor in front of the action in path integral. With fermions 
included, $\hbar/v^2$ appears explicitly in the Lagrangian (\ref{lagresc}),
in such a way that in the range of validity of the classical approximation
the gauge coupling of the $\si$ field is weak.

The temporal-gauge Hamiltonian density corresponding to (\ref{lagresc}) is
\bea
{\cal H}&=&\frac{1}{2}\left[E^2+|P|^2+\Pi^2+\left(\partial_x \si
          -  \frac{g\sqrt\hbar}{v\sqrt\pi} A\right)^2
    + |D_x\phi|^2\right]
  +{1\over 4}(|\phi|^2-1)^2\nonumber\\
   &&{}+{\cal Y}\left(\phi e^{-2iv\sqrt{\pi/\hbar}\si}
              + \phi^* e^{2iv\sqrt{\pi/\hbar}\si}\right),
\label{eq:H_inv}\eea
where $A$ is the spatial component of the gauge potential, whereas $E$, $P$ 
and 
$\Pi$ are the canonical conjugate momenta of $A$, $\phi$ and $\sigma$, 
respectively. In the following we solve numerically the equations of motion
obtained from this Hamiltonian. It is easy to verify that the time evolution
described by ${\cal H}$ preserves the local Gauss' constraints
\beq \partial_xE-g\sqrt{\hbar\over{\pi v^2}}\Pi-2ig(P\phi-P^*\phi^*)=0 
\label{gauss}\eeq

While the Hamiltonian (\ref{eq:H_inv}) is best suited for numerical description
of real-time evolution, the static properties of the model are made more
transparent by eliminating the $\sigma$ degree of freedom.
To this end, we perform the gauge transformation $\sigma\rightarrow 0$,
$A\rightarrow A+v{\sqrt\pi\over\sqrt\hbar g}\sigma'$ and
$\alpha\rightarrow\alpha+2v\sqrt{\pi\over\hbar}\sigma$,
where $\alpha={\rm Arg}\phi$.
We also solve the constraint (\ref{gauss}) for $\Pi$ and obtain for
the remaining degrees of freedom
\bea
{\cal H}&=&{1\over2}(E^2+P_\rho^2+{P_\al^2\over\rho^2}+{{\pi v^2}\over
{g^2\hbar}}(E'-2gP_\al)^2)
+{1\over2}\rho^2(\al'-2gA)^2+{1\over2}{\rho'}^2\nonumber\\
&&{}+{g^2\hbar\over2\pi v^2}A^2 +{1\over4}(\rho^2-1)^2+2{\cal Y}\rho\cos(\al),
\label{eq:Hamiltonian}
\eea
where $P_\rho$ and $P_\alpha$ are the radial and angular conjugate momenta
of the scalar field, respectively.
We see that the fermions induce a photon mass $g\sqrt{\hbar/\pi v^2}$.
In the classical regime, $\hbar/v^2\ll 1$, this mass is small compared to $2g$
the photon mass induced by the Higgs mechanism.
This form of the Hamiltonian is also used in the following for generating
the canonical ensemble of initial configurations at finite temperature.

It is important to understand how the fermions of the original
formulation appear in the bosonized version.
First, the baryon number in the $\si=0$ gauge is minus the Chern-Simons
number:
\beq
B=-{g\over\pi}\oint dxA\ .
\eeq
Keeping this in mind, we can analyze the vacuum structure of the theory.
In the fermion-less Abelian Higgs model, there is an exact degeneracy
of vacua labelled by the winding number of the scalar field. In these
vacua, related by topologically nontrivial gauge transformations,
the Chern-Simons number is equal to the winding number. 
Not so in the presence of fermions: the minima of the energy correspond
to different values of an observable quantity, the fermion number. Hence,
these minima are no longer related by a gauge transformation. Moreover,
these states have different energies.

To see how this comes about,
consider the minimum of the energy (\ref{eq:Hamiltonian}) for 
a macroscopically small winding number $n\ll L$, where $L$ is the
spatial size of the system.
Minimizing the static part of (\ref{eq:Hamiltonian})
with respect to $A$, we find
\beq
A={1\over 2g}{1\over 1+{\hbar\over 4\pi v^2\rho^2}}\alpha'.
\label{eq:Aalphaprime}
\eeq
In the case ${\cal Y}=0$ we take an ansatz of a constant $\rho$.
Using (\ref{eq:Aalphaprime}), we determine
that the constant $\al'={2\pi n\over L}$ minimizes the static energy, 
and therefore we have the relation
\beq
n=-[1+{\hbar\over 4\pi v^2\rho^2}]B
\eeq
where $n$ is the winding number of the field $\alpha$. We see that close to
the classical limit minima of the energy approximately correspond to
integer values of the fermion number. If $\rho$ is constant,
so is $\alpha'=2\pi n/L$,
whereas for the constant magnitude of the scalar field one finds
$\rho^2=v^2+{\cal O}((n/L)^2)$. Hence, the scalar self-coupling term in the 
potential approaches its absolute minimum in the infinite-volume limit.
We then obtain for the vacuum energy
\beq
{\cal E}(B)/L=\hbar\pi{B^2\over L^2}(1+{\hbar\over4\pi v^2})
+{\cal O}\left({B\over L}\right)^4.
\label{enomass}\eeq

In the general ${\cal Y}\neq 0$ case $\alpha'$ is no longer constant for a
minimal-energy winding solution. 
For simplicity let us consider the limit of the gauge and Yukawa couplings
being small compared to the scalar self-coupling: $g\ll 1$, ${\cal Y}\ll 1$.
Then $\rho$ does not deviate significantly from 1. In the
classical limit $\hbar/v^2\ll 1$ we again have that the baryon
number is equal to the winding number of $\alpha$. Consequently one
expects that the fermions will appear in the spectrum as the solitons
of the field $\alpha$. This indeed is the case. Using (\ref{eq:Aalphaprime}),
we obtain the static equation for $\alpha$:
\beq
\al''+2{\cal Y}(4\pi {v^2\over\hbar} + 1)\sin\al=0
\label{eq:sineg}
\eeq
This is the sine-Gordon equation,
which possesses soliton solutions. The mass of the fermion is equal to
the energy of the (one-winding) solution\footnote{The mass of 
the fermion in the Lagrangian eq.(\ref{eq:Lagrangian}) is
given by the relation $M_f=\hbar yv$. This however does not contradict the
bosonized result eq.(\ref{eq:fermionmass})
since the relation between the Yukawa
coupling $y$ in eq.(\ref{eq:lagr}) and the coupling $Y$ in
eq.(\ref{eq:Lagrangian}) is nonlinear. The standard bosonization
procedure leads to $Y\propto y^2$ (taking account of the normal
ordering of the exponential of $\si$), 
which is indeed consistent with eq.(\ref{eq:fermionmass}).}
\beq
M_f=8\sqrt{2{\cal Y}\over1+4\pi v^2/\hbar}.
\label{eq:fermionmass}
\eeq
Similarly, there exist multi-soliton solutions for any $n$. For a vanishingly 
small fermion density $n/L$ the $n$-fermion state has the energy $nM_f$.

At larger ${\cal Y}$ the winding solutions corresponding to fermion
excitations look somewhat differently.
It is not favourable energetically to keep $\rho$ spatially
constant. Instead in the region where the phase $\alpha$ varies
between $-\pi$ and $\pi$, the radial field $\rho$ is significantly
smaller than its value in the vacuum. This suppresses the contribution
of the kinetic energy of the $\alpha$ field to the energy.
At very large ${\cal Y}$ when
the mass of $\alpha$ is larger than the mass of $\rho$ this is obvious
since it is the only way to keep down the energetic cost of 
the winding configuration. 

Having discussed the static properties of the model, we can now identify the
relevant 
dimensional scales of the problem. This identification is important for
the numerical
study presented in the following. Consider the relevant length scales first.
These are the fermion (soliton) size of the order 
$\sqrt{\hbar/(8\pi{\cal Y}v^2)}$ and the sphaleron size, of order one (the 
sphaleron configuration is discussed in some detail in the next section).
The system size should be far above, and the spatial discretization (the lattice
spacing) far below any of these scales. Next, the relevant time scales are
the inverse frequencies ${\cal O}(2g)^{-1}$, ${\cal O}(1)$, and 
${\cal O}(2{\cal Y}^{-1/2})$ for the gauge, radial scalar, and angular scalar
modes, respectively. The time integration step should be chosen well below
any of these scales. Finally, consider the relevant energy scales. 
The sphaleron energy is of order 1. If we wish to be in the range of validity 
of the sphaleron approximation, we must insist on the inverse temperature 
$\beta\gg 1$. On the other hand, for the classical approximation to make sense,
the temperature 
must be well above the Higgs (${\cal O}(\hbar/v^2)$), the photon
(${\cal O}(2\hbar g/v^2)$), and the $\alpha$-meson 
(${\cal O}(2\hbar \sqrt{2{\cal Y}}/v^2)$) masses. For light fermions, an 
additional condition relating the temperature and the system size follows from
(\ref{enomass}).
Knowing that the fermion coupling to the gauge field is small in the
classical limit $\hbar/v^2\ll 1$, we can compare this energy to the
exact energy for $B$ free massless fermions, $\hbar\pi B(B+1)/L$,
whose relative deviation from (\ref{enomass}) is $1/B$. The latter is
small if a typical value of $B$ is large. At a finite temperature
$1/\beta$ this condition is achieved if $\pi\hbar\beta/v^2\ll L$.

In this work, we use, in view of these considerations,
the following choices for the various parameters:
gauge coupling $g=\sqrt{2.5}$,
Yukawa coupling $0\le{\cal Y}\le0.080$,
the loop expansion parameter $\hbar/v^2=0.05$,
inverse temperature $13\le\beta\le15$.
We performed simulations with the values of $a=0.25$ and $a=0.125$ and 
found no measurable lattice spacing dependence of the transition rate. 
Similarly, we varied the system length between 250 and 625, and observed no 
finite size effects.

\section{The sphaleron}

As we know, transitions can occur between the vacua of different winding
number. Of interest for the rate of these transitions is the
energy of the sphaleron configuration (the lowest energy barrier separating
the vacua).

For the massless fermion case (${\cal Y}=0$), Roberge has shown that
the sphaleron configuration is identical to the one in the no-fermion case
if the fermion density $n$ is macroscopically small \cite{roberge94}.

For the massive case ${\cal Y}\neq 0$ the sphaleron solution is not known 
analytically.
Instead, we determined the sphaleron energy,
applying the extremization method \cite{dumaw,sijs,vbaal}.
The idea is to look for static solutions of the classical equations of motion
(including sphaleron configurations) by minimizing
the sum of the
squares of the right hand sides of the equations of motion for the fields:
\beq
{\cal V}\equiv\sum \left({\partial V\over\partial\varphi_i}\right)^2.
\label{extrem}\eeq
The sum extends over all the degrees of freedom $\varphi_i$ where $V$
is the potential part of $\cal H$.
The minimization is done using a simple relaxation procedure, {\it i.e.,}
integrating relaxation equations 
$\partial_\tau\varphi_i=-\partial_{\varphi_i}{\cal V}$. In the following, 
we use the 
same procedure, with ${\cal V}$ replaced by $V$, to measure the integer part of
Chern-Simons number.

We used the method as described to determine the height of the sphaleron
barrier separating the minima with winding numbers 0 and 1, relative to the
absolute minimum of $V$.

\begin{figure}[ht]
\centerline{\psfig{file=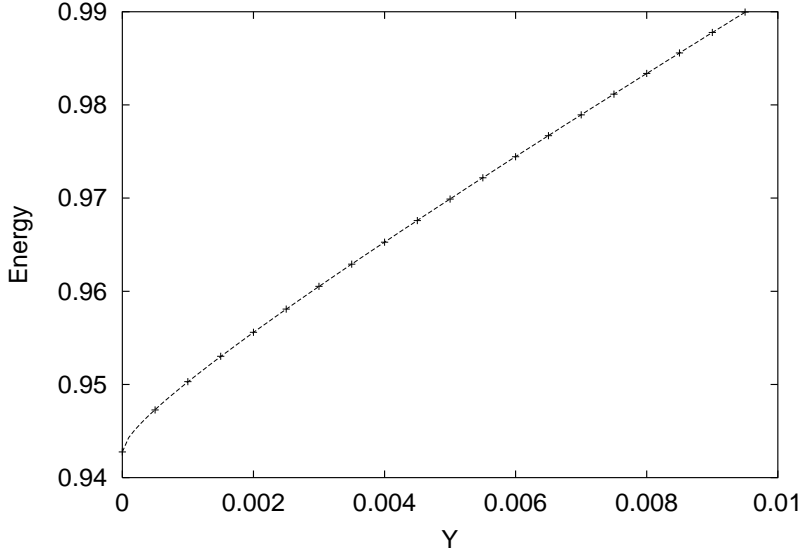,width=11cm}}
\caption{The energy of the sphaleron configuration relative to the vacuum 
energy. The pluses are the values obtained by the extremization method.
The solid curve denotes the least-squares fit to the form 
$a + b\sqrt{\cal Y} + c{\cal Y}$.}
\label{fig:sphalener}
\end{figure}

The results are displayed in Figure~\ref{fig:sphalener}.

As is clear from the figure, the sphaleron energy is well approximated
by the relation 
\beq
E_{sph}(Y)=E_0 + b \sqrt{\cal Y} + c {\cal Y}
\label{eq:Esph_Y}
\eeq
where a least squares fit yields
$E_0 = 0.942762 \pm 0.000015$,
$b = 0.11919 \pm 0.00056$,
$c = 3.744  \pm 0.005$.
As expected, $E_0$ coincides with the sphaleron energy for massless fermions,
which, in turn, is equal to the sphaleron energy in the absence of fermions.

\begin{figure}[ht]
\centerline{\psfig{file=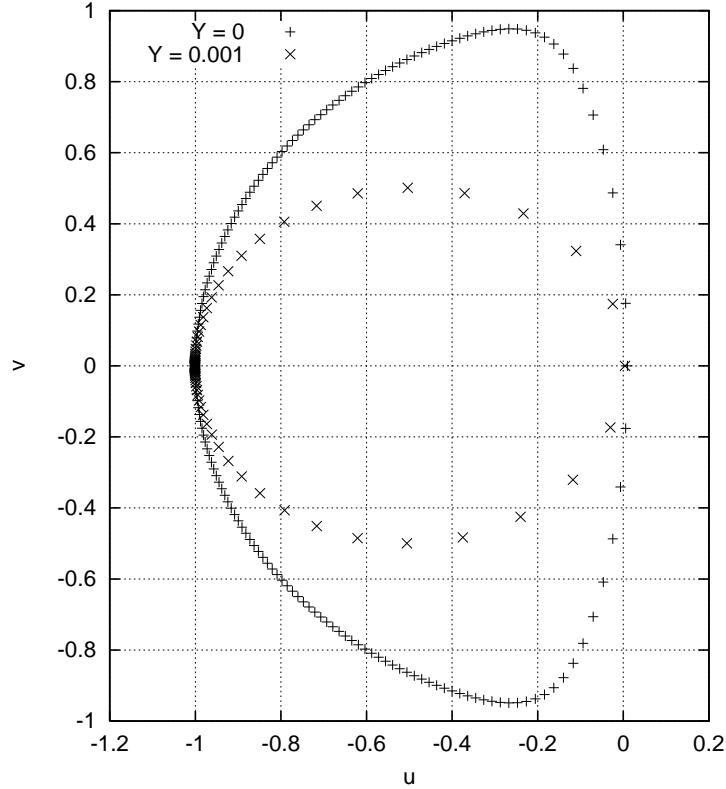,width=10cm}}
\caption{The real and
imaginary parts of the scalar field ($u$ and $v$) in the sphaleron
configuration, for ${\cal Y}=0$ and for ${\cal Y}=0.001$.}
\label{fig:sphalconfig}
\end{figure}

Figure~\ref{fig:sphalconfig} displays the values of the real and
imaginary parts of the scalar field ($u$ and $v$) for two configurations,
for ${\cal Y}=0$ and ${\cal Y}=0.001$.
What is clearly visible is that for ${\cal Y}=0$ the fields passes through 
zero,
and then winds slowly back, through the minimum $u^2+v^2=1$, to the
other side of the Mexican hat.
For ${\cal Y}=0.001$, the field passes with equal speed over the Mexican hat,
but immediately bends toward the minimum at $u=-1$, $v=0$.

The dependence displayed in (\ref{eq:Esph_Y}) can be given a qualitative
explanation.
The sphaleron energy measures the difference between the energy
of the sphaleron configuration and the vacuum energy.
The latter corresponds to the absolute minimum of the potential,
which occurs, for small positive $\cal Y$, for a field value close to
$-v$.
For ${\cal Y}=0$, the sphaleron configuration reduces to the fermionless
case
\beq
\Phi(x)=i\tanh(x/\sqrt{2})\exp(i\pi x/L).
\eeq

Corrections to the sphaleron energy at a finite $\cal Y$ can be estimated as
follows. First, the phase factor $\exp(i\pi x/L)$ is affected strongly
for any finite value of $\cal Y$.
This comes about because, the scalar field must revert
to its vacuum value in a finite interval, centered at $x=0$
to avoid an extensive ($\propto L$) contribution to the sphaleron energy.
Much like in the soliton solution of (\ref{eq:sineg}), the size of this interval
is ${\cal O}(\hbar/8\pi{\cal Y}v^2)^{1/2}$, in the classical regime where 
$\hbar/v^2\ll 1$. Likewise, the corresponding correction to the sphaleron energy
is of the same order as $M_f$ given by (\ref{eq:fermionmass}). As $\cal Y$
grows and becomes ${\cal O}(\hbar/8\pi v^2)$, there occurs a crossover from
${\cal O}(\sqrt{\cal Y})$ to ${\cal O}({\cal Y})$ behavior of the sphaleron 
energy correction, given, in this case, by the spatial integral of the 
Yukawa term in the sphaleron background.

\section{The sphaleron transition rate}

In the present work, we will be interested in
temperatures that are
large enough so that quantum effects can be ignored, so that
the dynamics is essentially described by classical field theory.
Moreover, we will restrict our attention in this work to temperatures
that are small relative to the sphaleron energy, so that the
transition rate is suppressed by the factor $e^{-\beta E_{sph}}$.
Our choice of the gauge and Yukawa couplings, temperature, linear size,
and discretization in space and time is discussed at the end of Section 2.

We use the familiar technique to simulate sphaleron transitions \cite{fkp}.
Namely, we use a combination of Metropolis and heat-bath Monte-Carlo
algorithms to draw initial conditions for real-time evolution from the canonical
ensemble at temperature $1/\beta$, corresponding to the discretized version of
(\ref{eq:Hamiltonian}). Having generated an initial condition, we switch
to the Hamiltonian (\ref{eq:H_inv}), better suited for real-time evolution. In
doing so, we set initially $\sigma=0$, whereas $\Pi$ is determined by the Gauss'
law.

\begin{figure}[ht]
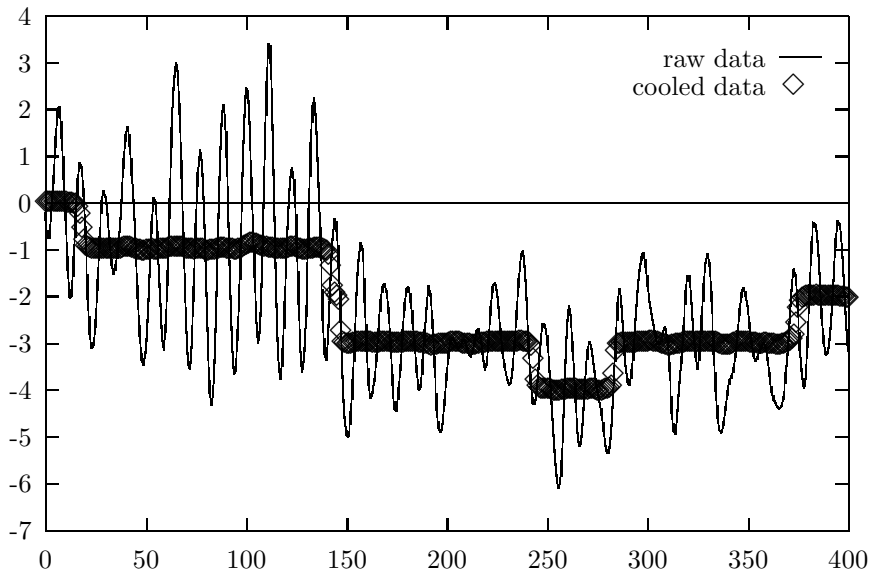

\input cool_uncool
\caption{A sample time history 
of the Chern-Simons number, measured before and after cooling.}
\label{fig:sample}
\end{figure}

However, the procedure for measuring the sphaleron rate requires modification
in the presence of fermions. The rate can no longer be defined, as it was
without fermions, as the diffusion constant per unit volume of Chern-Simons
number. Indeed, the energy is no longer a periodic function of Chern-Simons
number, hence the average squared topological charge can no longer grow
linearly with time.
We use an alternative measurement method, in which the transitions are counted
directly in a real-time simulation.
In order to easily identify sphaleron transitions,
the field configuration is subjected to relaxation (cooling). A configuration
can be thought to represent thermal fluctuations in the vicinity of an energy
minimum (a vacuum). The cooling eliminates thermal fluctuations. This is
done by solving, for every field $\varphi_i$, the relaxation equation
\beq
\partial_\tau\varphi_i=-\partial_{\varphi_i}{\cal V}.
\eeq
The algorithm is essentially the same
as the one employed to determine the sphaleron configurations, but
now using the regular static potential $V$ instead of $\cal V$ given by 
(\ref{extrem}). 
The resulting cooled configuration has
an approximately integer Chern-Simons number, and transitions can be easily
counted, as illustrated in Figure \ref{fig:sample}.

We performed three series of simulations.
In the first series we measured the sphaleron transition rate
in the Abelian Higgs model without the fermions, where the rate is already
known from earlier work \cite{kp,fkp}. The goal here was to compare the
rate obtained by counting transitions to the one found as the diffusion
constant of Chern-Simons number. The reliability of the former method depends
on the frequency of transitions. If the system size is too large or the 
temperature is too high, the counting method is not accurate because individual
transitions cannot be resolved. We verified that the agreement between the two
methods was excellent for all the combinations of sizes and temperatures
reported here.

In the second series of measurements we determined the transition rate for
massless (${\cal Y}=0$) fermions.
Table~\ref{tab:nofermmassless} summarizes the temperature dependence of
the rate without
and with massless fermions.

\begin{table}[ht]
\centerline{\begin{tabular}{|lrr|} \hline
$\beta$ & $\Gamma$ (no fermions) & $\Gamma$ (${\cal Y}=0$) \\
13 & $(16.2\pm 0.9)\times 10^{-5}$ & $(16.1\pm 1)\times 10^{-5}$ \\
14 & $(7.06\pm 0.4)\times 10^{-5}$ & $(6.68\pm 0.34)\times 10^{-5}$ \\
15 & $(3.06\pm 0.2)\times 10^{-5}$ & $(3.05\pm 0.21)\times 10^{-5}$ \\ \hline
\end{tabular}}
\caption{A comparative table of transition rate without fermions and with
fermions at zero Yukawa coupling.}
\label{tab:nofermmassless}
\end{table}

As is clearly visible, there is no measurable effect of the massless fermions
on the rate within the error bars.

In our final series of simulations
we investigated the transition rate in the presence
of massive fermions.
Specifically, the sphaleron transition rate was measured using the
cooling/counting method at a fixed values of $\beta$,
namely for $\beta=13,$ 14 and 15.
We considered values of $\cal Y$ between 0 and $0.080$.
\begin{figure}[ht]
\centerline{\psfig{file=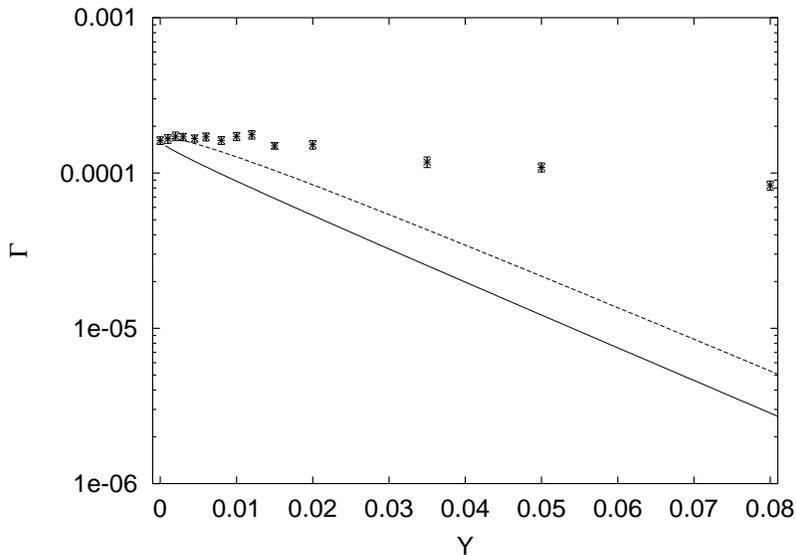,width=11cm}}
\caption{The sphaleron transition rate as a function of Yukawa coupling at 
$\beta=13$. The solid curve is $R_u$, the right-hand side of (\ref{eq:Ru}).
The dashed curve is $R$ as given by (\ref{eq:R}).}
\label{fig:beta13}
\end{figure}
\begin{figure}[ht]
\centerline{\psfig{file=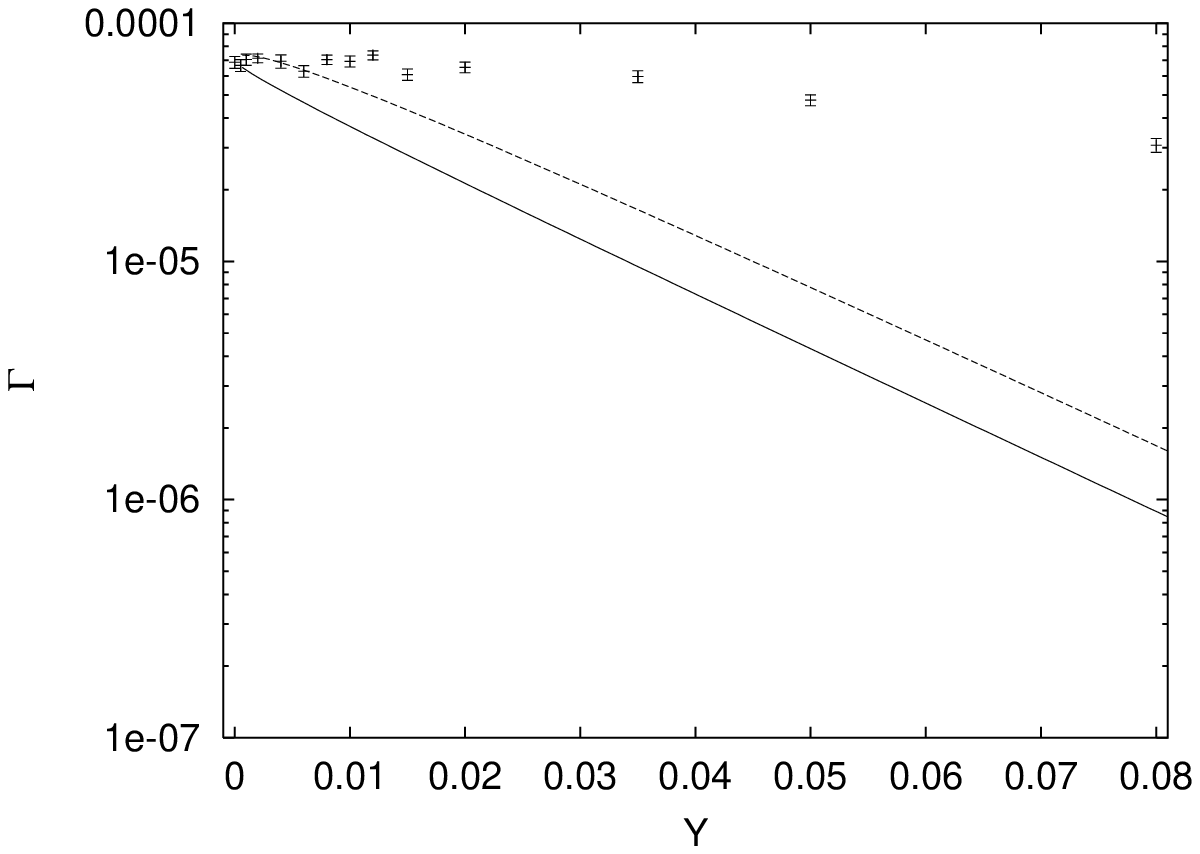,width=11cm}}
\caption{The sphaleron transition rate as a function of Yukawa coupling at $\beta=14$. The solid curve is $R_u$, the right-hand side of (\ref{eq:Ru}).
The dashed curve is $R$ as given by (\ref{eq:R}).}
\label{fig:beta14}
\end{figure}
\begin{figure}[ht]
\centerline{\psfig{file=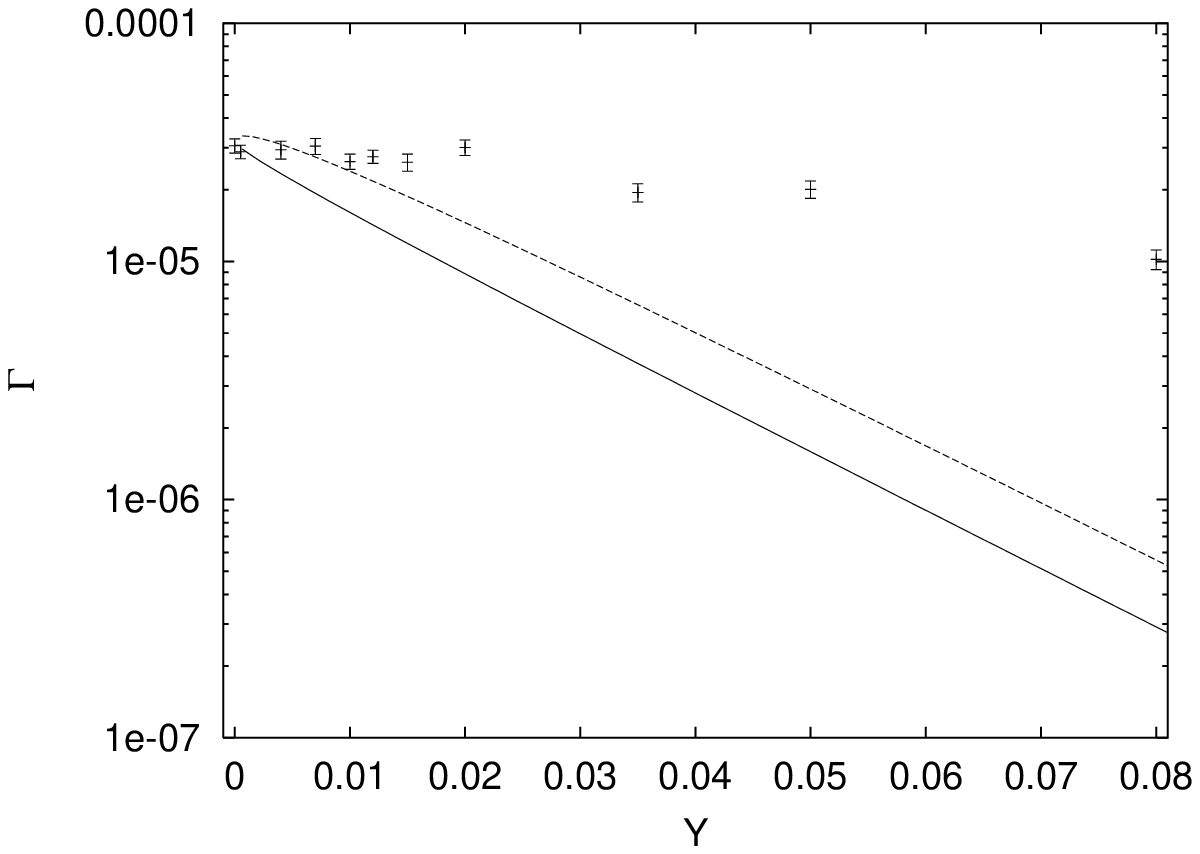,width=11cm}}
\caption{The sphaleron transition rate as a function of Yukawa coupling at $\beta=15$. The solid curve is $R_u$, the right-hand side of (\ref{eq:Ru}).
The dashed curve is $R$ as given by (\ref{eq:R}).}
\label{fig:beta15}
\end{figure}

The results are plotted in Figures \ref{fig:beta13}, \ref{fig:beta14}
and \ref{fig:beta15}.  As can be seen from the figures, the transition
rate, approximately constant as a function of $\cal Y$ for ${\cal Y}<0.02$,
drops off for ${\cal Y}>0.02$, the rate at ${\cal Y}=0.080$
being a factor 2 ($\beta=13$) to 3 ($\beta=15$) smaller than at ${\cal Y}=0$.
A dropoff of the rate as a function of $\cal Y$ is exactly what we expect,
as the sphaleron energy increases with $\cal Y$ according to (\ref{eq:Esph_Y}).
However, this rate dependence on $\cal Y$ is much weaker than
predicted by the sphaleron approximation, which would predict an exponential
dropoff (solid line).
We discuss this result in more detail in the following section.

\section{Discussion}

We investigated the effect of dynamical fermions on the sphaleron transition
rate, combining bosonization and the classical approximation. As discussed
in Section 2, the resulting theory retains important qualitative features
expected from the presence of dynamical fermions. This theory is also 
interesting in its own right, since it helps elucidate the role of additional
(other than the gauge field and the Higgs scalar) degrees of freedom in 
sphaleron transitions. Let us then summarize the lessons learned.

First of all, comparing the rate in the
presence of massless fermions to the rate without fermions,
we see that, within the error there is essentially effect of including the fermions is a factor that
is independent of the temperature for the range of temperatures
considered ($12<\beta<16$). In other words, the inclusion of
fermions does not alter the functional dependence on the temperature.
Such behavior agrees very well with what one would expect in the sphaleron 
approximation for the rate, which, in the absence of fermions, gives
\beq
\Gamma=\kappa\,T^{2/3}\left({E_{sph}\over T}\right)^{7/6}e^{-E_{sph}/T},
\label{eq:Ru}
\eeq
where $\kappa$ is a numerical constant.
In this formula the exponential is the Boltzmann factor of the sphaleron
configuration. As we saw in Section 3, massless fermions do not change
the sphaleron energy. The temperature dependence of the non-exponential
pre-factor results from the existence of zero modes in the sphaleron background
which are not present in the background of a vacuum configuration. Massless
fermions do give rise to new zero-modes, absent in the no-fermion case. In our
model, the new zero-mode is the $\alpha$-meson. However, this mode is not
associated with the sphaleron only, it exists both in the sphaleron and the 
vacuum backgrounds. Hence it cannot change the temperature dependence of the
pre-factor.

More interesting is the dependence of the rate on the Yukawa coupling.
As one can see from figures \ref{fig:beta13}, \ref{fig:beta14} and
\ref{fig:beta15}, the rate is approximately constant for $0<{\cal Y}<0.02$,
edging down slowly in the range $0.02<{\cal Y}<0.08$.

We are not aware of, nor do we attempt here, a rigorous determination of the
transition rate in the sphaleron approximation in presence of dynamical 
fermions.
However, in a first attempt to understand its dependence on the Yukawa
coupling,
one might expect that a reasonable first approximation to
the rate would be one that based on the rate formula (\ref{eq:Ru})
for the fermionless case.
The solid curve ($R_u$) in figures \ref{fig:beta13}, \ref{fig:beta14} and
\ref{fig:beta15} denotes the right-hand side of (\ref{eq:Ru}),
taking for $E_{sph}$ the $\cal Y$ dependent value given by (\ref{eq:Esph_Y}).
It is obvious that the exponential decay as a function of $\cal Y$ thus
predicted is inconsistent with the numerical data.

While we will not resolve this discrepancy in the present work,
we would like to point out two possible effects of the fermions on the rate.

\begin{figure}[ht]
\centerline{\psfig{file=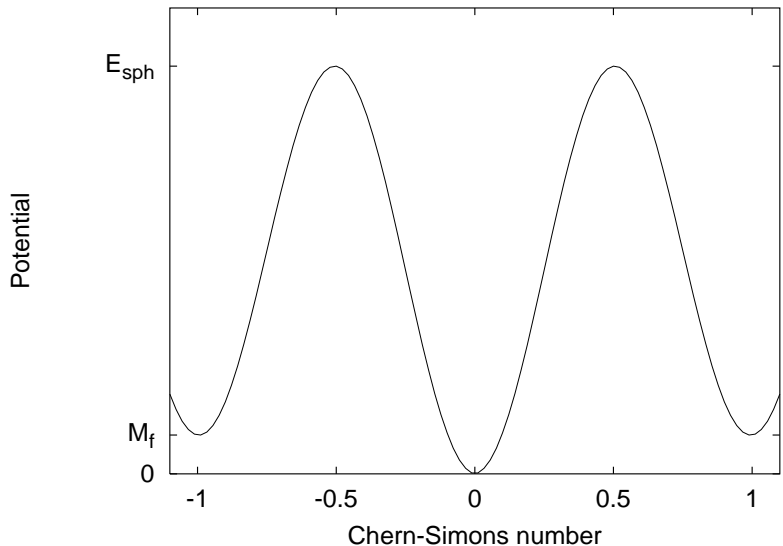,width=11cm}}
\label{fig:fermionenergy}
\caption{Schematic representation of the effective potential as a function
of Chern-Simons number.}
\end{figure}

The first one is related to the fact that the presence of fermions
lifts the degeneracy of the vacuum as a function of Chern-Simons number
to the energy of the created (anti)fermion, as is indicated
schematically in Figure \ref{fig:fermionenergy}.
In the sphaleron approximation the rate is controlled by the sphaleron
energy, or rather, the difference between the sphaleron energy and
the vacuum energy.
This would be applicable for the rate from the $N_{\rm CS}=0$ 
to the $N_{\rm CS}\approx\pm1$ state.
However, the corresponding Boltzmann factor for the reverse transition
is different (larger), due to the energy difference $M_f$ 
({\it cf} (\ref{eq:fermionmass})), by a factor
$e^{\beta M_f}$.
The same conclusion is reached by applying detailed balance at temperature
$T$.

To compare the sphaleron approximation with the measured rate,
we should apply an appropriate averaging over the various types
of transitions, that include creation as well as annihilation of
(anti)fermions.
To this effect, let us consider the following simplified model.
Our one-dimensional system can be divided into $g$ intervals of the order
of the soliton size.
Each interval can be in any of three states:
empty, occupied by a soliton (fermion), or occupied by an antisoliton (but
not both), each with energy $M_f$.
Transitions can take place from the empty state to the soliton
or antisoliton state with probability per unit time $R_u$,
while the reverse probability is $R_d$, by detailed balance equal
to $R_u e^{\beta M_f}$.
Direct transitions between soliton and antisoliton state are not possible.
The total transition rate will be equal to the $g$ times the
(Boltzmann-)averaged transition rate per interval.
The latter yields $(2R_u+2R_de^{-\beta M_f})/(1+2e^{-\beta M_f})
=4R_u/(1+2e^{-\beta M_f})$
(the denominator comes from normalization).
The total transition rate is simply $g$ times larger.
One finds for the total rate
\beq
R={4g \over 1+2e^{-\beta M_f}}R_u.
\label{eq:R}
\eeq
As $M_f$ is proportional to $\sqrt Y$, we thus find a non-trivial
$\sqrt{\cal Y}$-dependent exponential correction.
It is to be expected that a similar correction is present
in the formula for the effective sphaleron transition rate.

The rate $R$ is plotted for comparison with our numerical data
in the Figures \ref{fig:beta13} through \ref{fig:beta15}.
It is obvious that for ${\cal Y}>0.02$ $R$ is still in
disagreement with our numerical results.
A somewhat better agreement
between the data and $R$ at small values of $\cal Y$ is possibly
coincidental, since the description in terms of localized
non-overlapping solitons only applies at large $\cal Y$, such that
${\rm exp}\,(-\beta{\cal Y})$ is a small number.
We conclude that the
near independence of the transition rate on $\cal Y$ cannot be
explained away by the statistical distribution of fermions.

Another possible explanation for the very weak dependence of the rate
on $\cal Y$ is as follows.
In the zero fermion mass case one degree of freedom, the phase
of the scalar field, is a zero mode, and we have a corresponding
Goldstone boson.
For ${\cal Y}>0$, the degeneracy is broken, and the Goldstone boson acquires
a mass.
However, for very small values of $\cal Y$, the symmetry remains approximate,
and all values of the phase angle remain almost equally occupied.
In particular, we can expect a high fermion density.
This means that the Gaussian approximation around the field minimum
($\phi\approx-1$) will not be a good one, and one has to apply
an appropriate treatment of the Sine-Gordon model instead.
We intend to address this point in the future.

\section{Acknowledgements}
A.~K. and R.~P. wish to acknowledge financial support from the Portuguese
Funda\c c\~ao para a Ci\^encia e a Tecnologia, under grants
CERN/S/FAE/1177/97 and CERN/P/FIS/1203/98.


\section{References}

\begin{thebibliography}{xx}

\baselineskip = 18pt

\bibitem{smaa}
G.~Aarts and J.~Smit, Nucl.~Phys.~B {\bf 511}, 451 (1998).

\bibitem{grs}
D.Yu.~Grigorev, V.A.~Rubakov, and M.E.~Shaposhnikov, Phys.~Lett.~B {\bf 216}, 
172 (1989).

\bibitem{bf}
A.I.~Bochkarev and P.~de~Forcrand, Phys.~Rev.~D {\bf 44}, 519 (1991).

\bibitem{kp}
A.~Krasnitz and R.~Potting, Phys.~Lett.~B {\bf 318}, 492 (1993).

\bibitem{fkp}
P.~de Forcrand, A.~Krasnitz, and R.~Potting, Phys.~Rev.~D {\bf 50}, 6054 (1994).

\bibitem{st}
W.H.~Tang, J.~Smit, Nucl.~Phys.~B {\bf 540}, 437 (1999).

\bibitem{aaps}
J.~Ambj{\o}rn, T.~Askgaard, H.~Porter, and M.E.~Shaposhnikov,
Phys.~Lett.~B {\bf 244}, 479 (1990); \\ Nucl.~Phys.~B{\bf 353}, 346 (1991).

\bibitem{ak31}
J.~Ambj{\o}rn and A.~Krasnitz, Phys.~Lett.~B {\bf 362}, 97 (1995); \\
Nucl.~Phys.~B{\bf 506}, 387 (1997).

\bibitem{gm}
G.D.~Moore, Nucl.~Phys.~B {\bf 480}, 657 (1996); \\ Phys.~Lett.~B {\bf 412}, 
359 (1997); \\ Phys.~Lett.~B {\bf 439}, 357 (1998); \\ Phys.~Rev.~D {\bf 59},
014503 (1999).

\bibitem{gmt}
G.D.~Moore and N.~Turok, Phys.~Rev.~D {\bf 56}, 6533 (1997).

\bibitem{gr} T.~M.~Gould and I.~Z.~Rothstein, Phys.~Rev.~D {\bf48}, 5917
(1993).

\bibitem{roberge90} A.~Roberge, Phys.~Rev.~D {\bf41}, 2605 (1990).

\bibitem{roberge94} A.~Roberge, Phys.~Rev.~D {\bf49}, R1689 (1994).

\bibitem{dumaw}A.~Duncan and R.~Mawhinney, Phys.~Lett.~B {\bf 282}, 423 (1992).

\bibitem{sijs}A. van der Sijs, Phys.~Lett.~B {\bf 294}, 391 (1992).

\bibitem{vbaal} M.~G.~Perez, P.~van Baal, Nucl.~Phys.~B {\bf429}, 451
(1994); Nucl.~Phys.~B~{\bf468}, 277 (1996).

\bibitem{moore} G.D.~Moore, Phys.~Rev.~D {\bf53}, 5906 (1996). 

\bibitem{mitya} D.~Diakonov, M.~Polyakov, P.~Sieber, J.~Schaldach and K.~Goeke,
Phys.~Rev.~D {\bf49}, 6864 (1994).

\bibitem{kuntz} G.~Nolte and J.~Kuntz, Phys.~Rev.~D {\bf48}, 5905 (1993).

\bibitem{smaaf} G.~Aarts and J.~Smit, "Real-time dynamics with fermions on a
lattice", hep-ph/9812413 (1998).

\bibitem{bochkarev} A.~I.~Bochkarev and  G.~G.~Tsitsishvili,
Phys.~Rev.~D~{\bf40}, 1378 (1989).

\end{thebibliography}
\end{document}